\documentclass[%
 reprint,
superscriptaddress,
 amsmath,amssymb,
 aps,
]{revtex4-1}

\usepackage{graphicx}
\usepackage{dcolumn}
\usepackage{bm}

\usepackage[version=3]{mhchem}

\newcommand{\Z}{\ce{ZrSiS}}
\newcommand{\Hf}{\ce{HfSiS}}
\newcommand{\C}{$^{\circ}\text{C}$}
\newcommand{\D}{$^{\circ}$}

\usepackage{upgreek}

\begin{document}


\title{Quantum oscillations of the magnetic torque in the nodal-line Dirac semimetal ZrSiS }

\author{F. Orbanić}
\affiliation{Department of Physics, Faculty of Science, University of Zagreb, Croatia}

\author{M. Novak}
\affiliation{Department of Physics, Faculty of Science, University of Zagreb, Croatia}

\author{Z.Glumac}
\affiliation{Josip Juraj Strossmayer University of Osijek, Croatia}

\author{A. McCollam}
\affiliation{High Field Magnet Laboratory, Radbound University, Nijmegen, the Netherlands}

\author{L.Tang}
\affiliation{High Field Magnet Laboratory, Radbound University, Nijmegen, the Netherlands}

\author{I. Kokanović}
 \altaffiliation[]{Corresponding author: kivan@phy.hr}
\affiliation{Department of Physics, Faculty of Science, University of Zagreb, Croatia}




\date{\today}

\begin{abstract}
We report a study of quantum oscillations (QO) in the magnetic torque of the nodal-line Dirac semimetal \Z\ in the magnetic fields up to 35 T and the temperature range from 40 K down to 2 K, enabling high resolution mapping of the Fermi surface (FS) topology in the $k_z=\pi$ (Z-R-A) plane of the first Brillouin zone (FBZ). It is found that the oscillatory part of the measured magnetic torque signal consists of low frequency (LF) contributions (frequencies up to 1000 T) and high frequency (HF) contributions (several clusters of frequencies from 7-22 kT). Increased resolution and angle-resolved measurements allow us to show that the high oscillation frequencies originate from magnetic breakdown (MB) orbits involving clusters of individual $\alpha$ hole and $\beta$ electron pockets from the diamond shaped FS in the Z-R-A plane. Analyzing the HF oscillations we unequivocally shown that the QO frequency from the dog-bone shaped Fermi pocket ($\beta$ pocket) amounts $\beta=591(15)$ T. Our findings suggest that most of the frequencies in the LF part of QO can also be explained by MB orbits when intraband tunneling in the dog-bone shaped $\beta$ electron pocket is taken into account. Our results give a new understanding of the novel properties of the FS of the nodal-line Dirac semimetal \Z\ and sister compounds.       
\end{abstract}

\pacs{Valid PACS appear here}
\maketitle


\section{\label{sec:level1} Introduction}

In recent years the discovery of Dirac and Weyl type excitations in the low-energy band dispersion of the Dirac and Weyl semimetals (DSM and WSM) represents a major breakthrough in condensed matter physics \cite{1,2,3,4,5,6,7,8}. Due to their unique band topology, they show different exotic electronic properties of technological and fundamental interest. Dirac semimetals can become Weyl semimetals or can be driven to other exotic topological phases such as topological insulators and topological superconductors by breaking certain symmetries which determine the band topology of the material \cite{9,10,11,12,13,14,15,16,17}. Unlike Dirac or Weyl semimetals, where there are discrete touching points between the valence and conduction band in the first, in a nodal-line semimetal there are symmetry protected band degeneracies which form lines [closed loops or open lines in the first Brillouin zone (FBZ)]. If the material posses time reversal and inversion symmetries, these crossing lines will be fourfold degenerate (analogous to Dirac points in a DSM) and we are talking about a nodal-line Dirac semimetal (NLDSM). Due to their unique band topology, effects like charge order, magnetism, and superconductivity are theoretically predicted to occur in NLDSM materials \cite{19,20,21,22,23,24,25,26,27}.     

\Z\ is a member of the $\rm{MX'X''}$ group of compounds \cite{MX'X''}, where $\rm{M}$ is a metal (Zr, Hf, Ta, Nb), $\rm{X'}$ is a +2 valence state of Si, Ge, As and $\rm{X''}$ belongs to the chalcogen group. \Z\ and sister compounds have recently gained a lot of attention due to the symmetry protected crossing of the conduction and valence bands which results in the NLDSM phase. In the system with no spin-rotation symmetry [namely a system with spin-orbit coupling (SOC)] additional non-symmorphic crystal symmetries (glide planes and a screw axis) are required to protect the NLDSM phase \cite{Oxide, Fang}, which is the case in \Z\ and sister compounds \cite{Chen, Schoop, Topp, Hu_2}. In \Z\ there are symmetry protected nodal lines running parallel to the $k_z$ direction in the FBZ (X-R and M-A directions) located deeper in the valence band. There is another set of nodal lines forming a cage-like structure in the FBZ which are closer to the Fermi energy. The later nodal lines are not symmetry protected and thus susceptible to a small gap opening due to SOC in \Z. While most of the transition-semimetal materials studied so far have linear band dispersion up to a few hundred meV from the Dirac node, in \Z\, this energy range is as high as 2 eV in some regions of the FBZ. Thus, the primary criterion to observe exotic properties related to nodal-line Dirac fermions, that the Fermi energy of the semimetal should remain within the linear dispersion region, is fulfilled in \Z. The main goal of this work is a detailed study of the \Z\ Fermi surface (FS) using cantilever torque magnetometry for high quality home-grown crystals.

Recent studies of the FS morphology in \Z\ by angle-resolved photoemission spectroscopy (ARPES) and quantum oscillations (QO) measurements confirmed the nature and position of three-dimensional (3D) electron and quasi two-dimensional (2D) hole pockets, and another two small pockets with quantum limits at around 10 and 32 T \cite{Chen,Schoop,Neupane,Hosen,Marcin,Hu,Zhang_Gao,Sankar,Ali,Z_1,Z_2}. A calculated 3D representation of \Z\ FS can be seen in \cite{Z_1, Z_2}. Since in this work we analyze QO in the magnetic torque signal (dHvA oscillations) for magnetic fields in directions near to the crystalline $c$ axis (or near to the $k_z$ direction in momentum space), we show that the most relevant part of the FS for an explanation of the measurement data is a cross section of the FS at $k_z=\pi$ (Z-R-A plane) in the FBZ, schematically shown later in Fig. \ref{fig_orbits}. It consists of four $\beta$ electron and four $\alpha$ hole pockets in a diamond shape configuration separated by a small gap (10-20 meV) due to the small SOC. 

Very recently, a high magnetic field study of \Z\ semimetal has revealed interesting MB orbits and significant mass enhancement of the quasiparticles residing near the nodal loop, whereas, in the sister compound \Hf, an effect of Klein tunneling in momentum space between adjacent electron and hole pockets in the top Z-R-A plane of the FBZ has been reported \cite{Z_1, Z_2, H_1}.             

The FS of \Z\ and sister compounds is still not fully understood. In this paper, in an effort to fully understand the FS of \Z\ in the top Z-R-A plane of the FBZ and how it's shape affects electron dynamics in high magnetic field, we performed magnetic QO measurements using the highly sensitive piezo-resistive cantilever torque technique at low temperature in magnetic fields up to 35 T and mapped the experimental frequencies to the calculated FS \cite{Z_1,Z_2}. From the FFT analysis of the high-resolution torque magnetometry data we obtained a number of high oscillation frequencies in the range from 7 kT up to 12 kT as well as their harmonics. We confirmed that these arise from the MB orbit clusters of the individual $\alpha$ hole [oscillation frequency of 241(4) T] and $\beta$ electron [oscillation frequency of 591(15) T] pockets. In \cite{Z_1}, Pezzini \textit{et al}. observed almost identical high oscillation frequencies from 7 up to 12 kT, but no harmonics around 22 kT. Most of the LF FFT spectra of QO in \Z\ (frequencies from $ca$ 100 to 1000 T) is also explained by MB orbits in the Z-R-A plane of the FBZ. The clearly observed LF  set of QOs, which were not all identified in the previous study \cite{Z_1} can be explained as an MB effect with a linear combination of  $\alpha$ and $\beta$ pockets and an additional $\beta/2$ contribution that corresponds to a putative intra-pocket MB tunneling in the $\beta$ electron pocket. The $\gamma=415(8)$T pocket could be explained within the experimental error bar as an effect of MB with $\beta/2$ taken into account. There is also a "figure of eight" orbit with a frequency of 348(7) T. Moreover, by analyzing the separation of the harmonics of HF MB orbits, we confirmed the area of individual $\alpha$ and $\beta$ orbits of the FS in the top Z-R-A plane of the BZ. It is found that the dog-bone shaped Fermi electron pocket has an oscillation frequency of 591(15) T, which is different from the result in \cite{Z_2}.     

\begin{figure}[t]
\centering
\includegraphics[width=0.5\textwidth]{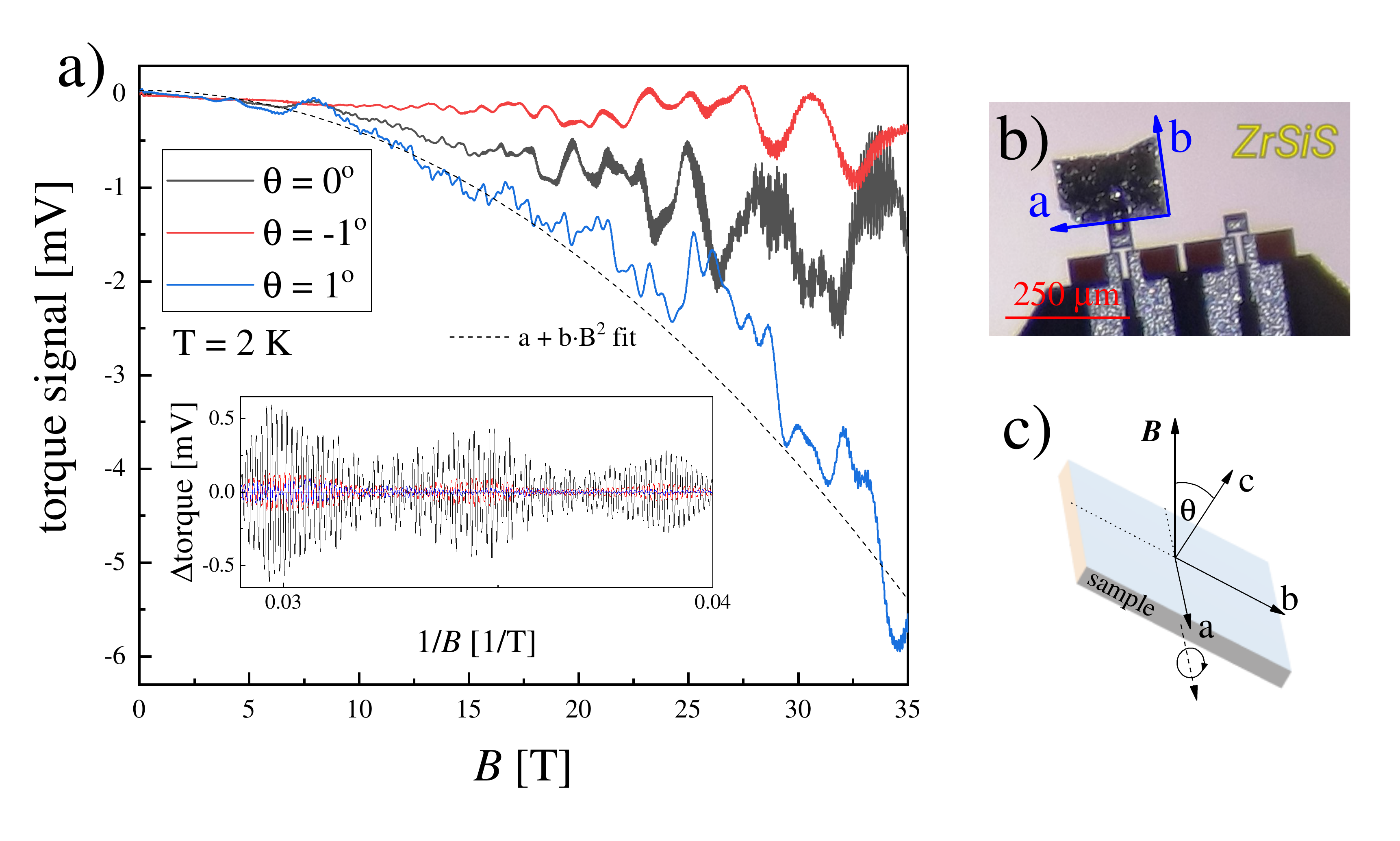}
\caption{a) The magnetic torque signals from a \Z\ crystal at 2 K, for $\uptheta=-1$\D, $\uptheta=0$\D, $\uptheta=1$\D. The torque signal consists of a $B^2$ dependent background with QO contributions superimposed. Dashed line shows $a+bB^2$ fit to the measured magnetic torque signal. The inset to FIG. 1 a) shows an enlarged view of isolated HF QO as a function of $1/B$ for $\uptheta=-1$\D, $0$\D\ and $1$\D. b) Image of a typical piezoresistive chip with \Z\ crystal glued on the cantilever. c) A schematic picture of sample and magnetic field configuration. By rotation of the sample around the crystalline $a$ axis the angle $\uptheta$ between the direction of magnetic field and the crystalline $c$ axis was varied. }
\label{fig_1}
\end{figure}

\section{\label{sec:level1} Experimental results}

Single crystals of \Z\ were grown by standard chemical vapour transport \cite{synthesis}. Their excellent quality is shown by their low-\textit{T} in-plane resistivity of only 0.1 $\upmu\Omega \rm{cm}$. The \Z\ crystals were characterized by x-ray diffraction which confirms a tetragonal PbFCl-like compound structure with \textit{P4/nmm} space group. The temperature dependence of the resistivity $\rho_{ab}$ of the crystals at zero field shows metallic behavior with a residual resistivity ratio ($RRR=\rho_{300\rm{K}}/\rho_{1.8\rm{K}}$) of around 80.

The magnetic torque signal of the \Z\ single crystal was measured in a \ce{^4He} cryostat measurement system with a single-axis rotator option, using commercially-available piezoresistive cantilevers (SEIKO-PRC120), in steady fields up to 35 T and in the temperature range from 2 to 40 K \cite{lever}. The direction of the applied magnetic field is determined by simultaneously measuring the Hall voltage of a Hall probe. The offset between the maximum of the Hall probe signal and the $c$ axis of the crystal is expected to be smaller than 1\D\ and is neglected in the analysis of the data. The magnetic torque signal was measured for various angles $\uptheta$ between the crystalline $c$ axis and the direction of the magnetic field, Figs. \ref{fig_1} (b) and (c). The measurement data were taken during up and down magnetic field sweeps from 0 to 35 T. 

The magnetic torque signals of a \Z\ crystal in magnetic fields up to 35 T at the temperature 2 K and angles  $\uptheta=-1$\D, $0$\D\ and $1$\D, shown in Fig. \ref{fig_1} (a), highlights the main experimental observations: (i) the magnetic torque signal is a superposition of a $B^2$ dependent contribution from crystal magnetism and multi-frequency QO, (ii) above the MB threshold field of around 13 T, the torque signals reveal a series of HF QO, which are strongly suppressed by small angle misalignment between the direction of magnetic field and the $c$ crystalline axis [the inset to FIG. 1 a)]. The inset to Fig. \ref{fig_1} (a) shows an enlarged view of isolated HF QO contribution as a function of $1/B$ for angles $\uptheta=-1$\D, $0$\D\ and $1$\D. An image of the \Z\ single crystal glued on the end of the piezoresistive lever is shown in Fig. \ref{fig_1} (b), whereas Fig. \ref{fig_1} (c) displays a schematic view of the axis of rotation and the angle $\uptheta$ between the direction of magnetic field and the $c$ crystalline axis. The magnetic torque signal of the crystal was also measured at temperatures of 4.2, 10, 20 and 40 K. Different frequency contributions in the oscillatory part of the measured magnetic torque data have been distinguished by fast Fourier transform (FFT) analysis.  

\begin{figure}[t]
\centering
\includegraphics[width=0.45\textwidth]{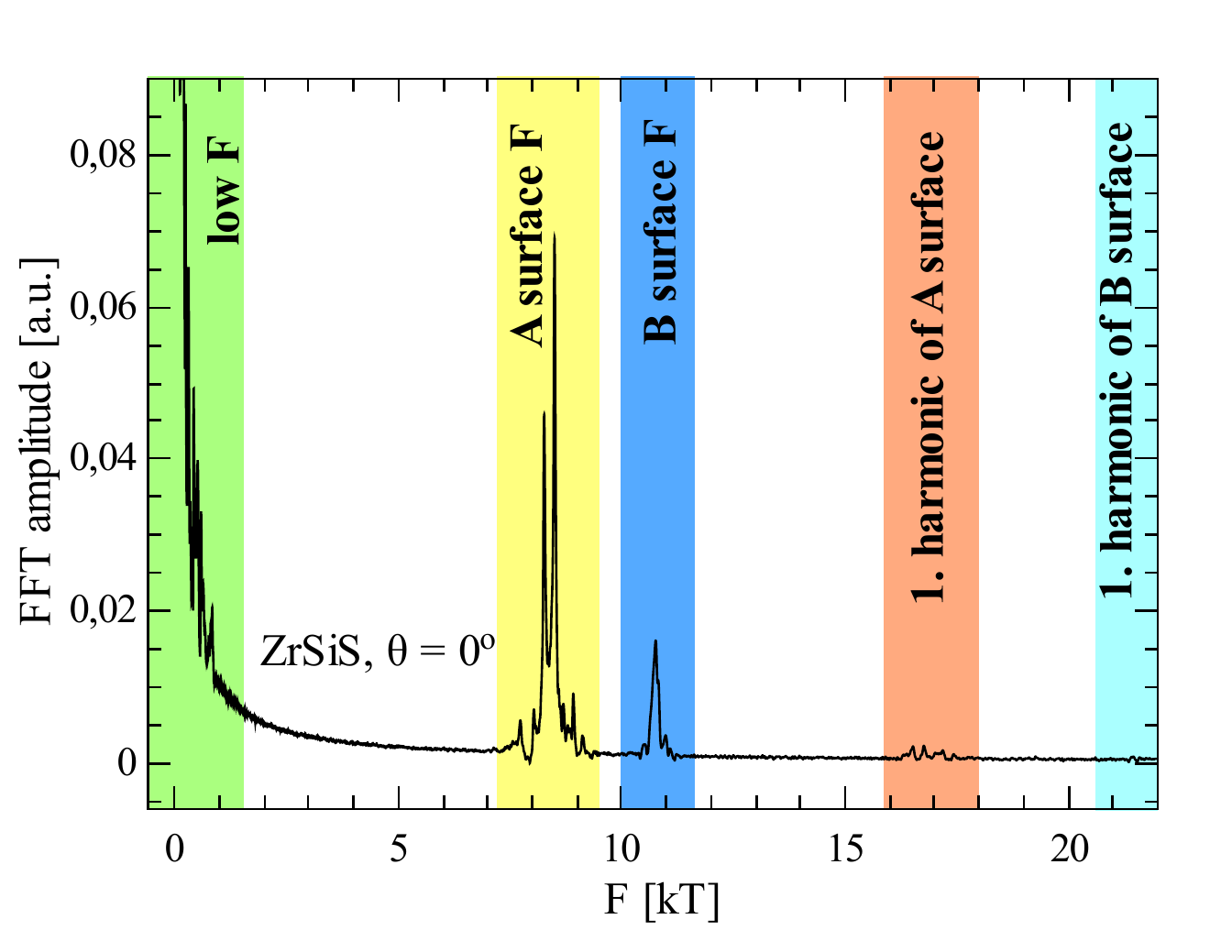}
\caption{The FFT spectrum of QO in the measured magnetic torque of \Z\ crystal for $\uptheta=0$\D\ at 2 K in the magnetic field range from 0 up to 35 T. The FFT frequency spectrum consists of LF contribution (frequencies up to 1 kT) and several clusters of HF contributions (frequencies from 7-22 kT). The high oscillation frequencies are attributed to different electron MB orbits as shown in Fig. \ref{fig_orbits}. The first harmonic of the MB $\rm{A}+\rm{n}\alpha$ (yellow bar) and $\rm{B}+\rm{n}\alpha$ (blue bar) orbits are clearly visible and highlighted in brown and light blue, respectively. }
\label{fig_FFT_spectrum}
\end{figure}

The FFT spectrum of the measured magnetic torque signal of the \Z\ crystal at 2K for $\uptheta=0$\D and the magnetic field range from 0 to 35 T is shown in Fig. \ref{fig_FFT_spectrum}. From the peaks in the FFT spectrum we identify the QO frequencies $F_k$, which are related to the extremal cross-sectional areas $A_k$ of the FS and the plane normal to the magnetic field direction via the Onsager relation $F_k=\frac{\phi_0}{2\pi^2}A_k$ with $\phi_0$ being flux quantum. As can be seen in Fig. \ref{fig_FFT_spectrum}, the FFT spectrum consists of the LF contribution (frequencies up to 1 kT) highlighted in green and two clusters ($\rm{A}$ and $\rm{B}$) of high frequencies and their harmonics highlighted in different colors (frequencies from 7-22 kT). 

\begin{figure}[hbtp]
\centering
\includegraphics[width=0.4\textwidth]{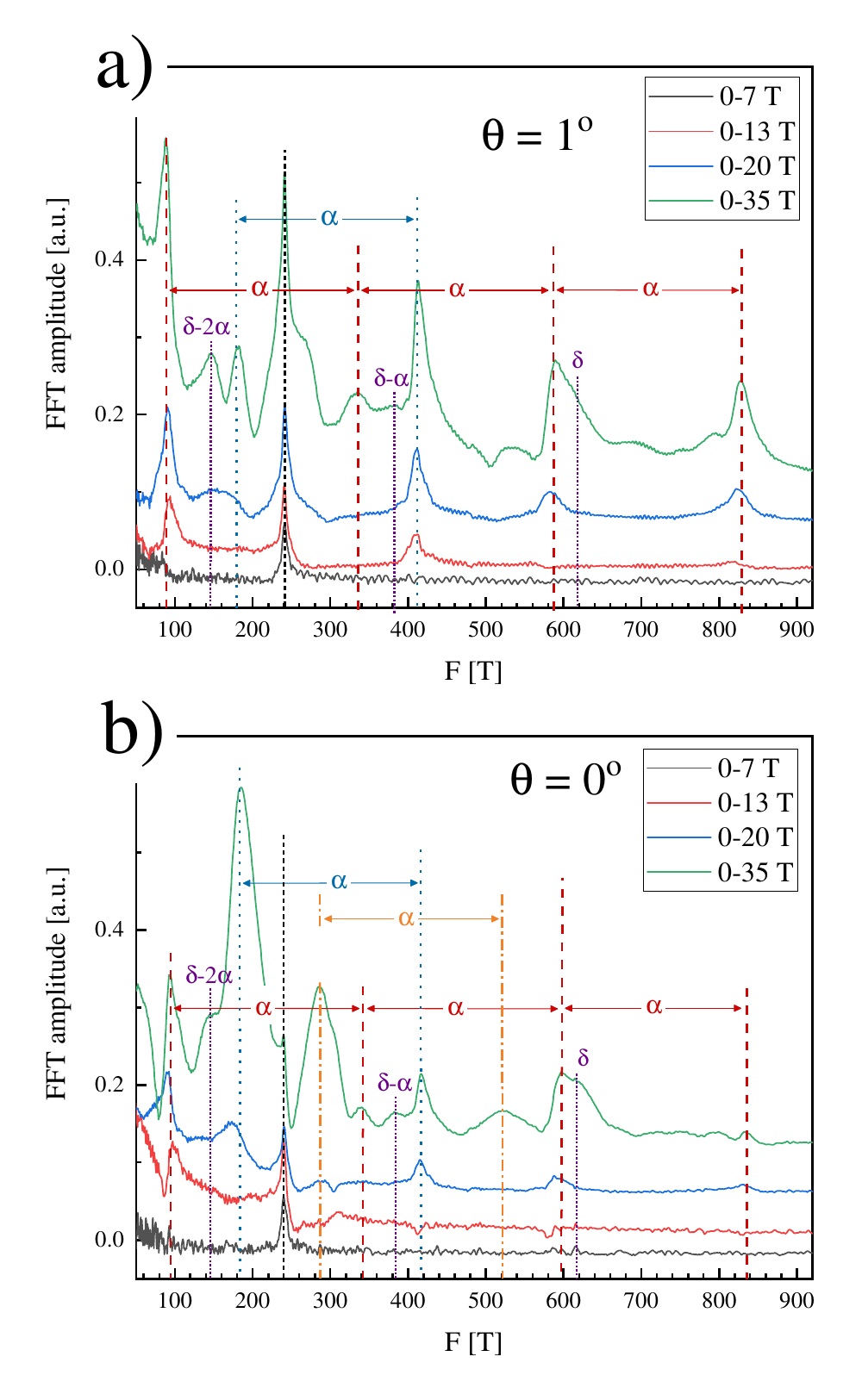}
\caption{The LF FFT spectrum of QO in the magnetic torque signal of \Z\ crystal at 2 K, obtained for the different magnetic field ranges and angles (a) $\uptheta=-1$\D and (b) $\uptheta=0$\D. Frequencies $\alpha=241(4)$ T and $\beta=591(15)$ T are contributions from $\alpha$ and $\beta$ pockets from the FS in the Z-R-A plane of the FBZ. Prominent high field peaks with frequencies $\gamma=415(8)$ T, $\eta=286(8)$ T and $180(5)$ T and other smaller peaks could be ascribed to various MB orbits, as explained in Sec. \ref{sec:levelB}. It is interesting to notice that at high fields, a small change in angle $\uptheta$ leads to a strong change in FFT spectrum and several of the peaks are $\alpha$ apart, giving a strong indication of the MB effect.}
\label{fig_LF}
\end{figure}

The LF contributions to QO in the magnetic torque of the same \Z\ crystal at 2 K for angles $\uptheta=0$\D\ and $\uptheta=1$\D\, obtained from the measured magnetic torque data in different magnetic field ranges are shown in Figs \ref{fig_LF} (a) and (b), respectively. As can be seen from the LF FFT spectrum for the magnetic field range from 0 to 7 T, only one peak at 241(4) T is observed, which corresponds to the $\alpha$ hole pocket located at the vertex of the diamond-shaped FS in the top Z-R-A plane in the FBZ of \Z\ (Fig. \ref{fig_orbits}). For wider magnetic field ranges, the appearance of the new peaks in the LF FFT spectrum is strongly dependent on magnetic field angle tilted from the $c$ axis of the crystal, Fig. \ref{fig_LF}. Analyzing the FFT peaks for the magnetic field in the range from 0-35 T one can notice that the FFT spectra consist of peaks at 241(4) T ($\alpha$), 591(15) T ($\beta$), 415(8) T ($\gamma$), 286(8) T ($\eta$, for $\uptheta=0$\D) and peaks separated from these peaks by a multiple of $\alpha$, see Figs. \ref{fig_LF} (a) and (b). 

In the FFT spectra below 100 T we can clearly see frequencies at 8 and 22 T, which are already observed in several studies \cite{Chen,Schoop,Neupane,Hosen,Marcin,Hu,Zhang_Gao,Sankar,Ali,Z_1,Z_2,Z_Mario} and attributed to the different parts of \Z\ FS. These frequencies are not related to the MB effects discussed in this work and so for clarity the origin of the x axis is set at 50 T in the Fig \ref{fig_LF}. The amplitude of the most pronounced peak at 8 T is independent of the magnetic field range at which the FFT is performed and of angle $\uptheta$ (for $\uptheta=-1,0,1$\D). As such it is taken as a reference for normalization of the FFT spectra.    

The HF part of the FTT spectra of QO data in magnetic torque signal of the \Z\ crystal, for various temperatures listed in the main panel, is shown in Fig. \ref{fig_HF} a). Figure \ref{fig_HF} (a) is obtained by performing the FFT of an isolated HF contribution to QO shown in Fig. \ref{fig_HF} (b) (QO were isolated from the raw signal by subtracting the mean value of envelope curves of maxima and minima of HF QO). The HF FFT spectra consists of two main clusters of equidistant peaks separated by the $\alpha$ pocket frequency of 241(4) T. These are labeled as A and B MB orbits in FIG. \ref{fig_HF} a), their harmonics are shown in the inset to Fig. \ref{fig_HF} (a). Figure \ref{fig_HF} (c) shows the HF FFT spectrum for angles $\uptheta=-1$\D, $0$\D, $1$\D\ and reveals much stronger suppression of the $\rm{B+n}\alpha$ cluster by tilting the magnetic field direction with respect to the $c$ crystal axis than for the $\rm{A}+\rm{n}\alpha$ cluster. Our results confirm the results obtained in \cite{Z_1} but with an enhanced resolution so that the FS structure can be studied in more detail. 

\section{\label{sec:level3} Discussion}

The FS of \Z\ calculated using density functional theory (DFT) is presented in many papers \cite{Z_1,Z_2,Z_Mario}. It is an open FS in the $k_z$-direction (due to the quasi two-dimensional crystal structure of \Z) built from several electron and hole pockets. States from the extremal cross sections of the FS and plane normal to the direction of magnetic field are responsible for QO. It turns out that the most relevant cross section for explanation of experimental results at $\uptheta$ close to $0$\D\ is the diamond-shaped FS in the $k_z=\pi$ (Z-R-A) plane of the FBZ, see Fig. \ref{fig_orbits}. The Z-R-A plane of the FBZ consists of four electron ($\beta$) pockets and four hole ($\alpha$) pockets separated by a small gap (10-20 meV) due to the small SOC in \Z\ \cite{Z_1}. Theoretically predicted value of $\alpha$ and $\beta$ pocket frequencies in \Z\ are $\alpha=235$ T and $\beta=596$ T \cite{Z_1}. There are many experimental confirmations of the $\alpha$ frequency in \Z\ \cite{Marcin,Hu,Zhang_Gao,Sankar,Ali,Z_1,Z_2} while the $\beta$ frequency is, so far, seen only in few cases \cite{Z_1, Marcin} at magnetic fields higher than 10 T.

\subsection{\label{sec:level21} Magnetic breakdown}

Most of the experimental results presented here can be explained by the effect of MB. MB is an important quantum complement to the semiclassical Lifshitz and Onsager theory of metals in which tunneling of charge carriers between separated Fermi pockets in $k$-space is taken into account \cite{Shoenberg}. In the MB regime, charge carriers tunneling through a momentum gap in $k$-space between adjacent pockets leads to the observation of extremal MB orbits consisting of combinations of closed extremal orbits whose effective area can be much larger or smaller than the area of the individual Fermi pockets in the given plane of the FBZ. In the case of MB tunneling, an electron moves classically along the FS except in the close vicinity of the MB gap which acts like a two-channel scattering center. Thus, an initial wave of unitary amplitude entering a MB gap is separated into a transmitted wave with amplitude $A_{\text{t}}=iP^{1/2}$ and a reflected wave with amplitude $A_{\text{r}}=(1- P)^{1/2}$, where the tunneling probability, $P$, depends on the MB field, $B_{\text{MB}}$, according to $P=e^{-B_{\text{MB}}/B}$. The probability of an electron tunneling through the MB $k$-space gap ($k_{\text{g}}$) depends on the magnetic field $B$, $k_{\text{g}}$ and also on the local band-dispersion $E(k)$. An estimate of the $B_{\text{MB}}$ can be made using the formula $B_{\text{MB}}=\frac{\pi\hbar}{2e}\left(\frac{k_{\text{g}}^3}{a+b}\right)^{1/2}$, where $a$ and $b$ are the $k$-space radii of curvatures of the orbits on each side of the gap \cite{BMB}. Calculations show that variations in local FS curvature can easily change $B_{\rm{MB}}$ by a factor of 10 \cite{Obrien}. For multiple closed orbits all possible closed orbits are additive and the contribution from each orbit is multiplied by the MB reduction factor $R_{\text{MB}}=CA_{\text{t}}^{l_{\text{t}}}A_{\text{r}}^{l_{\text{r}}}$, where $l_{\text{t}}$ represents the number of MB points the orbit traverses by transmission, $l_{\text{r}}$ represents the number of MB points the orbit traverses by reflection, and $C$ is the weighting factor which depends on the symmetry of the orbit and represents the number of possible realizations of the given effective MB orbit. According to the direction of circulation of the electron momentum around individual orbits in momentum space, which build an effective MB orbit, the resulting QO frequency associated with the MB orbit will be the difference or sum of frequencies associated with individual orbits (different direction of circulation leads to the difference of frequencies and vice versa), Fig. \ref{fig_orbits} (b).

\begin{figure*}[hbtp]
\centering
\includegraphics[width=0.8\textwidth]{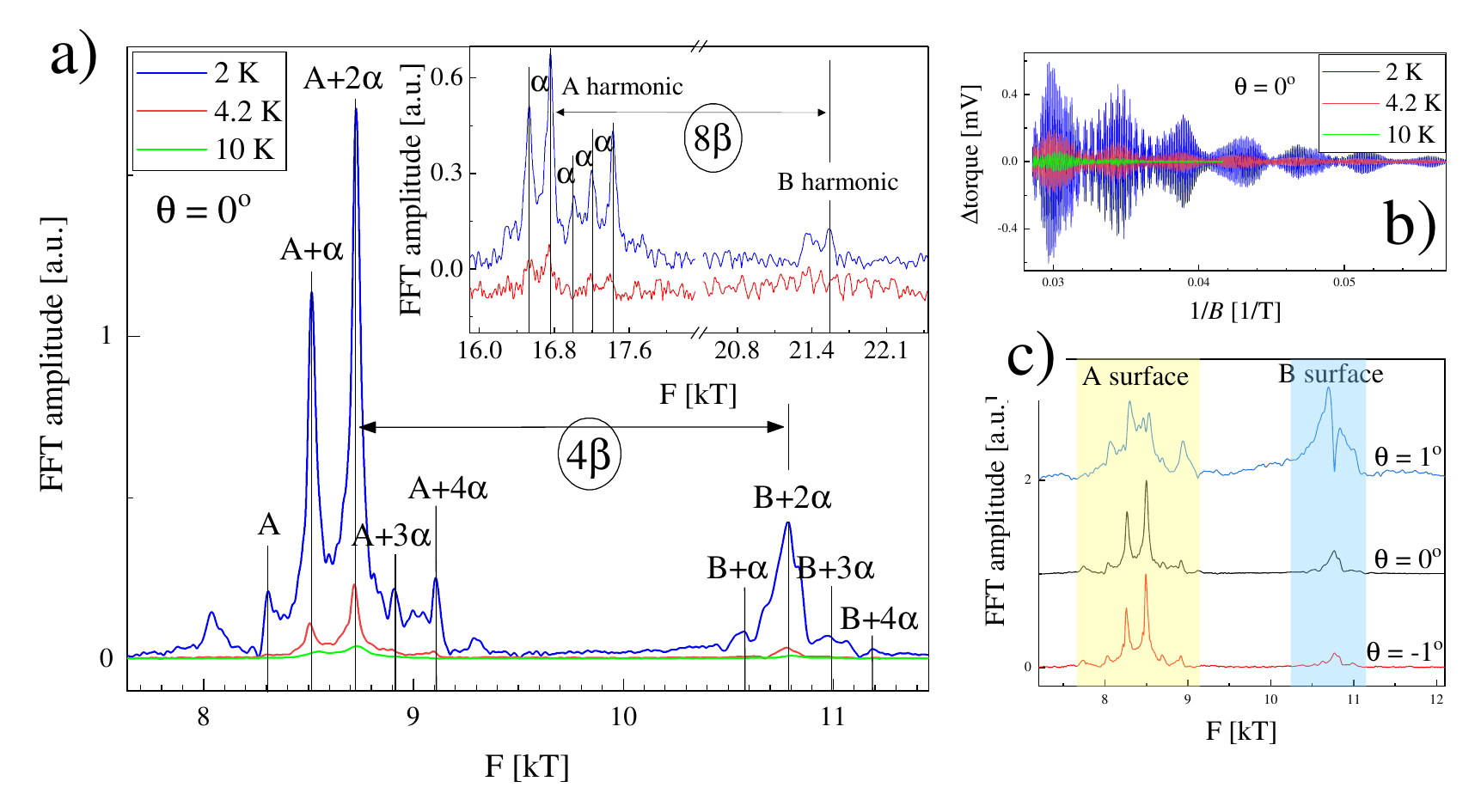}
\caption{a) The FFT of HF contribution to QO in the magnetic torque signal of \Z\ crystal measured at the different temperatures indicated in the main panel. The HF FFT spectrum consists of two main clusters of equidistantly separated peaks by the $\alpha$ hole pocket (A and B clusters, see Fig. \ref{fig_orbits}) and their harmonics [see the inset to Fig. \ref{fig_HF} (a)]. The calculated FFT spectra is explained by the effect of MB. According to the picture of MB orbits on the FS, A and B clusters of peaks and their harmonics should be separated by $4\beta$ and $8\beta$ respectively. Panel (b) highlights an enlarged view of the HF QO in the magnetic torque signal of \Z\ for the temperatures of 2, 4.2 and 10 K as a function of 1/B. c) FFT of HF contribution to QO in the magnetic torque signal of \Z\ for angles $\uptheta=-1$\D, $0$\D\ and $1$\D. Figure illustrates strong angle dependence of the MB orbits. }
\label{fig_HF}
\end{figure*}

\begin{table}[h]
\caption{Table of experimentally measured QO frequencies shown in the FIG. \ref{fig_LF}}
\begin{center}
\begin{tabular}{ |c|c||c|c| }
\hline
 Orbit & $\rm{F}_{exp}$ [T] & Orbit & $\rm{F}_{exp}$ [T]  \\ 
 \hline\hline 
  $\alpha$ &           241(4)                       &  {\footnotesize $\beta+\alpha$}   & 828(8)    \\
 \hline
  $\beta$  &           591(15)                       & {\footnotesize $\beta-\alpha$}  & 348(7)      \\
 \hline
 {\footnotesize $\gamma\approx\frac{3}{2}\beta-2\alpha$} & 415(8)&{\footnotesize $\beta-2\alpha$}&95(6) \\
 \hline
 {\footnotesize $\eta\approx\frac{\beta}{2}$}       & 286(8) &{\footnotesize $\gamma-\alpha$} & 180(5) \\
 \hline
 {\footnotesize $\delta\approx 2\eta$}            & 615(17)   & {\footnotesize $\eta+\alpha$}& 522(10) \\
 \hline
     {\footnotesize $\delta-\alpha$}    &    380(8)       & {\footnotesize $\delta-2\alpha$}  & 145(6)  \\
 \hline\hline
\end{tabular}
\end{center}
\label{tab}
\end{table}

\subsection{\label{sec:levelB} LF spectra}

In the LF FFT spectra in Figs \ref{fig_LF} (a) and (b), for magnetic fields larger than 20 T, a series of clearly resolved peaks are identified. Those are labeled as $\alpha$, $\beta$, $\gamma$ and $\eta$, and peaks that can be clearly matched with values $\beta-2\alpha$, $\beta-\alpha$, $\beta+\alpha$, $\gamma-\alpha$ and $\eta+\alpha$. The $\alpha$ and $\beta$ peaks correspond to magnetic oscillations of individual Fermi pockets ($\alpha$ hole petal pocket and $\beta$ electron dog-bone pocket in Fig. \ref{fig_orbits}) with frequencies of 241(4) T and 591(15) T, respectively. The claim that $\alpha$ and $\beta$ peaks correspond to the individual FS orbits in the top Z-R-A plane is discussed in more detail later when discussing the HF FFT spectra. One can notice that $\gamma$ and $\eta$ peaks at 415(8) T and 286(8) T are in an excellent numerical agreement with putative relations $\gamma\approx(\alpha+\beta)/2$ and $\eta\approx\beta/2$. The peak at $180(5)$ T could be seen as $\gamma-\alpha$ within the error bar. The peak $\delta$ with frequency near 600 T, which makes the $\beta$ peak broaden, can be attributed to the orbit $2\eta$. Associated peaks $\delta-\alpha$ and $\delta-2\alpha$ are also visible, Fig. \ref{fig_LF}. This suggests that most of the peaks, except $\alpha$ and $\beta$, can be associated with MB orbits. One can notice that, for the magnetic field range from 0 T to 35 T, there is a strong angle dependence of the $\alpha$, $\eta$ and $\gamma$ pockets as well as $\gamma-\alpha$, $\beta-\alpha$ and $\beta+\alpha$ pockets for the angle close to B parallel to the crystalline $c$ axis, Figs \ref{fig_LF} (a) and (b). Note that at $\uptheta=0$\D\ the signal from $\gamma-\alpha$ is large while the signals from $\alpha$ and $\gamma$ are suppressed, whereas the opposite is true for $\uptheta=1$\D. This is to be expected for orbits which are strongly affected by MB because there will be a higher proportion of MB orbits when the FFT is performed over a wider range of fields. 

At low magnetic fields, i.e., for fields less than the MB field, $B<B_{\rm{MB}}$, only $\alpha$ hole and $\beta$ electron pocket contribute to QO. For $B>B_{\rm{MB}}$ ($B_{\rm{MB}}$ may vary for different orbits) the MB-driven closed orbits appear at $\beta-2\alpha$, $\beta-\alpha$, $\beta+\alpha$, $\gamma$, $\gamma-\alpha$, $\eta+\alpha$. These are connected by $\alpha$ orbits so that Landau bands may develop. In the case of \Z\ both $\alpha$ and $\beta$ orbits contribute to the QO and the FFT spectrum of these oscillations becomes more complicated because the frequencies of $\alpha$ and $\beta$ are incommensurate, in general. Therefore, the frequency with which the $\beta$ Landau band cross the Fermi level (as a function of $1/B$) may be less or more than the $\beta$ frequency plus and minus integer multiples of the $\alpha$ frequency. The above picture of QO differs from that in the standard LK approach \cite{LK, A.A}. The principal difference is that in LK theory an external magnetic field does not change the energy spectrum of electrons and only determines the cross section of the FS. On the other hand, in the MB case the energy spectrum itself becomes a complex quasiperiodic function of the inverse magnetic field and strongly influences the QO spectrum \cite{A.A}. The QO frequency spectrum therefore becomes much more complex as MB gives rise to combined extremal orbits consisting of individual extremal orbits of the FS. 

Our further assumption is that there is an additional intraband tunneling in the $\beta$ pocket at the neck of the dog-bone shaped $\beta$ pocket, Fig. \ref{fig_orbits} (c). In this case an electron can orbit around nearly half of the $\beta$ pocket, leading to a broader $\eta\approx\beta/2$ frequency peak observed in the LF FFT spectra (Fig. \ref{fig_LF}). According to this picture, the frequency $2\eta$ should also be present. The peak associated with MB orbit $\delta \approx 2\eta$ makes the $\beta$ peak broaden, which can be seen in Figs \ref{fig_LF} (a) and (b). One possible combined MB orbit, in which an electron tunnels once through the neck of the $\beta$ pocket, shown in Fig. \ref{fig_orbits} (c), leads to the oscillation frequency $\frac{3}{2}\beta-2\alpha$. Because $\beta\approx 2.5\alpha$ in \Z\, this frequency amounts $\approx(\alpha+\beta)/2$, which can explain the $\gamma$ peak in the LF FFT spectra of \Z. The peak observed at 348(7) T matches the difference in frequencies of $\beta-\alpha$ demonstrating a "figure of eight" orbit, shown in Fig. \ref{fig_orbits} (b), that results from tunneling between adjacent $\beta$ electron and $\alpha$ hole pockets in the Z-R-A plane in the FBZ of \Z. Such a "figure of eight" orbit in the member of the same family of materials \Hf\ was reported for the first time in \cite{H_1} as a manifestation of Klein tunneling in momentum space \cite{Obrien}.    

All QO frequencies in the LF part of the FFT spectra shown in Fig. \ref{fig_LF} are summarized in Table \ref{tab}. 

O'Brien \textit{et al}. \cite{Obrien} showed that in the case of a type-II Weyl semimetal, the amplitude of the breakdown tunneling between the electron and hole Fermi pockets has a very strong dependence on the angle of the magnetic field with respect to to the axis of the Weyl cones. Thus, a strong variation of the peak amplitudes in LF FFT within 1\D\ around $\uptheta=0$\D, seen in Fig. \ref{fig_LF}, additionally confirms our conclusion that most of the peaks in LF FFT come from MB orbits. 

Further, it is well known that the intensity of each frequency peak in the FFT depends on the strength of disorder in the crystals, which suppresses the amplitude of oscillations. Thus, it depends also on the perimeter of the closed electron orbit. The peak broadening from disorder makes the $\beta=591(15)$ T pocket disappear in magnetic fields lower than 13 T as shown in the Figs. \ref{fig_LF} (a) and (b). This is because larger broadening makes faster oscillations (larger frequency) less visible. The disorder strength also affects the intensity of the other frequencies. \\

In principle the mixing frequencies $\beta\pm\alpha$, $\gamma\pm\alpha$ and their harmonics can also be produced by the effect of torque interaction \cite{torque int}, where the response of electrons to the induced oscillating magnetic moment is also taken into account. In large magnetic field the torque interaction can become so strong causing a distortion of the magnetization signal which leads to an enhancement of main frequency harmonics amplitude in FFT spectra. The effect of magnetic interaction increases with angle $\uptheta$. In our measured data no distortion of oscillating signal is observed so this scenario has been ruled out in our experiment.

\subsection{\label{sec:level2} HF spectra}

\begin{figure}[hbtp]
\centering
\includegraphics[width=0.47\textwidth]{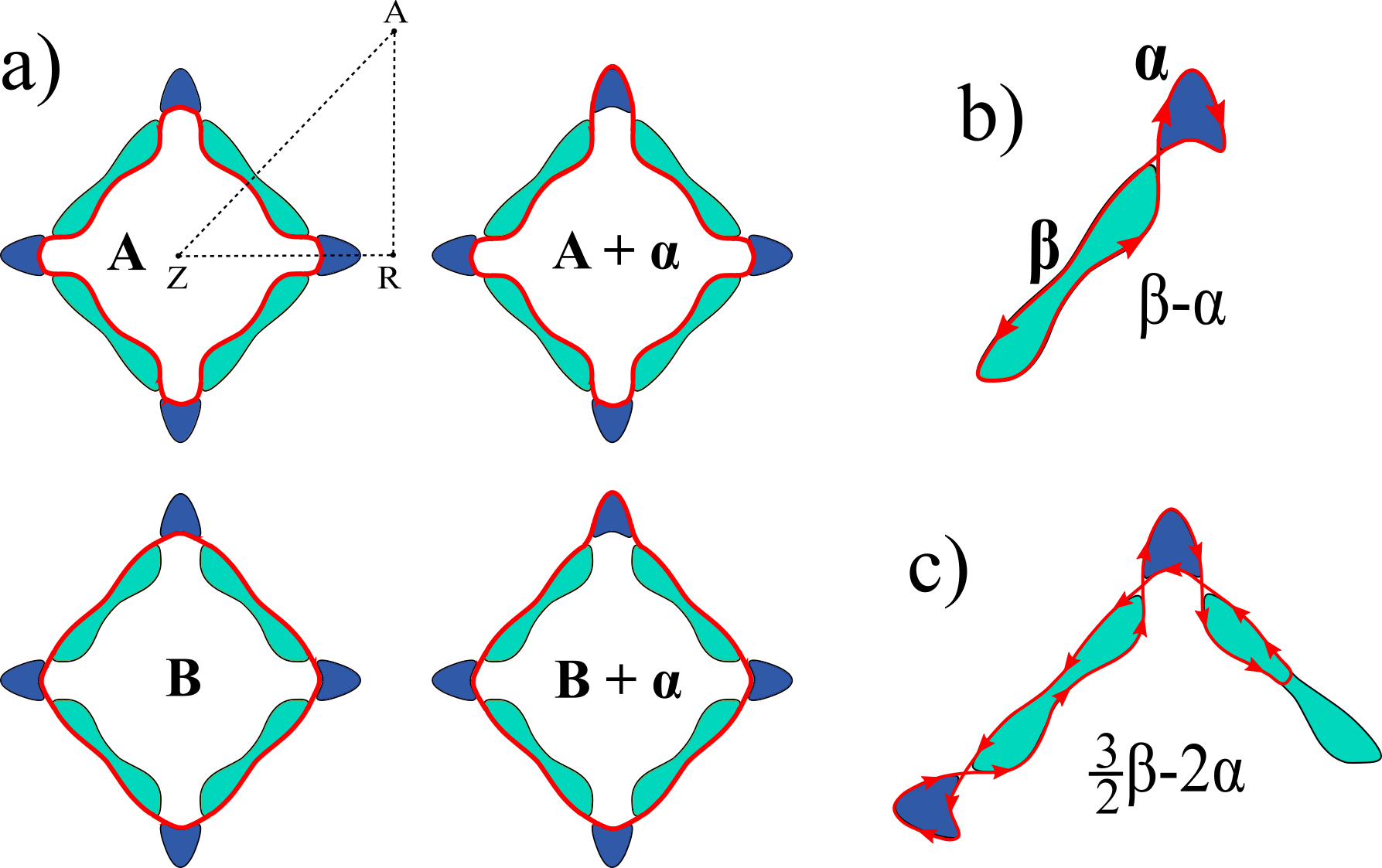}
\caption{Schematic representation of the FS of \Z\ in the Z-R-A plane of the FBZ, with MB orbits highlighted with red lines \cite{Z_1}. FS consist of four $\beta$ electron (light blue dog-bone shaped pocket) and four $\alpha$ hole pockets (dark blue pocket) in diamond shape configuration separated by a small gap due to the small SOC in \Z. (a) A red line indicates HF MB orbits where an electron encircles the entire FS in the Z-R-A plane. An electron can traverse the $\beta$ pocket over inner (A surface) or outer edge (B surface) and pick up between 0 to $\rm{n}$ $\alpha$ pockets. This explains the equidistantly separated HF FFT peaks by the $\alpha$ frequency inside $\rm{A}+\rm{n}\alpha$ and $\rm{B}+\rm{n}\alpha$ clusters. (b) The "figure of eight" MB orbit connecting one electron and one hole pocket resulting in $\beta-\alpha$ frequency of QO. It can be seen as an example of Klein tunneling in momentum space. (c) Example of MB orbit with tunneling through the neck of the dog-bone shaped $\beta$ pocket. The MB orbit shown explains the observed QO frequency $\gamma=415(8)$ T which can represented as a trajectory of the form $\frac{3}{2}\beta-2\alpha$. }
\label{fig_orbits}
\end{figure}

Next we will focus on the HF part of the FTT spectra of QO data in the magnetic torque signal of \Z\ crystal which is shown in Fig. \ref{fig_HF} (a) for various temperatures listed in the main panel. There are no individual orbits in the DFT-derived FS with areas that match the observed high frequencies in FFT spectra. Thus, the peaks of the $\rm{A}+\rm{n}\alpha$ and $\rm{B}+\rm{n}\alpha$ clusters in the HF FFT spectra correspond to MB orbits that encircle the entire diamond-shaped FS in the Z-R-A plane in the FBZ of \Z, see Fig. \ref{fig_orbits}. The $\rm{A}+\rm{n}\alpha$ and $\rm{B}+\rm{n}\alpha$ groups of peaks in the HF spectrum correspond to orbits in which an electron traverses $\beta$ pocket over the inner (A surface) or outer (B surface) edge. HF MB orbits can additionally include 0 to $\rm{n}$ $\alpha$ pockets on the path of an electron. This explains the equidistantly separated peaks by the $\alpha$ hole frequency of 241(4) T inside $\rm{A}+\rm{n}\alpha$ and $\rm{B}+\rm{n}\alpha$ clusters. According to this picture of MB electron orbits in momentum space, $\rm{A}+\rm{n}\alpha$ and $\rm{B}+\rm{n}\alpha$ clusters of peaks should be separated by $4\beta$ pockets, which is true in our case [for $\beta=591(15)$ T], see Fig. \ref{fig_HF} (a). As can be seen in the inset to Fig. \ref{fig_HF} (a), the first harmonics of the $\rm{A}+\rm{n}\alpha$ and $\rm{B}+\rm{n}\alpha$ clusters are clearly visible in the HF FFT spectrum. Further, according to this MB picture, harmonics of the two clusters should be separated by $8\beta$ pockets, which is again true in our case [for $\beta=591(15)$ T], see inset to Fig. \ref{fig_HF} (a). As well as in the $\rm{A}+\rm{n}\alpha$ cluster, individual peaks in the first harmonic of $\rm{A}+\rm{n}\alpha$ cluster are also separated by $\alpha$, not $2\alpha$ as one might naively expect. The first harmonic of the main frequency in QO comes from electrons which traverse the closed orbit in momentum space twice before they scatter. By performing a double orbit around the entire diamond shaped FS in the $k_z=\pi$ plane of the FBZ, an electron can again pick up from 0 to $\rm{n}$ additional $\alpha$ pockets. 

HF QO begin to appear at nearly 13 T. Therefore, above 13 T electrons gain enough energy to tunnel through at least eight gaps between $\alpha$ and $\beta$ pockets. One can assume that at this field the amplitudes of HF QO (amplitudes of HF FFT peaks) are dominantly determined by the weighting factor $C$ (introduced in Sec. \ref{sec:level21} on MB). MB orbit $\rm{A}+2\alpha$ ($\rm{B}+\rm{n}\alpha$) has the highest weighting factor, $C=6$ (2 of the 4 $\alpha$ pockets can be selected in ${4\choose 2}=6$ ways), so the peak with the highest amplitude in the $\rm{A}+\rm{n}\alpha$ ($\rm{B}+\rm{n}\alpha$) cluster of HF FFT is assigned to the $\rm{A}+2\alpha$ ($\rm{B}+2\alpha$) MB orbit.

HF MB orbits have a strong angle dependence, which can be seen in Fig. \ref{fig_HF} (c). Amplitude of $\rm{B}+n\alpha$ cluster of orbits is strongly affected by the change of angle $\uptheta$ suggesting that for some angles $\uptheta$ electrons have a preferred way of tunneling between $\alpha$ and $\beta$ pockets (going over inner or outer edge of the $\beta$ pocket). This effect can probably be associated with strong variation of peaks amplitude in LF FFT spectra within $\uptheta=0-1$ \D\, seen in Figs. \ref{fig_LF} (a) and (b). For some more quantitative conclusion about the angle dependence (within $\uptheta=0-1$ \D) of energy gap between $\alpha$ and $\beta$ pocket a more detailed experiment is needed (planned for the near future).

\section{\label{sec:level1} Conclusion}

It has been shown that the measurement of QO in the magnetic torque, by the piezoresistive cantilever method, provide a very precise probe for the FS of the top Z-R-A ($k_z=\pi$) plane of the FBZ of the NLDS \Z. The FS topology deduced from the magnetic  torque measurements is in very good agreement with DFT calculations, consisting of four $\alpha$ hole and four $\beta$ electron individual orbits which give rise to the diamond-shaped FS in the top  Z-R-A  plane of the FBZ. 
  
We showed that A and B clusters of peaks in the HF FFT part correspond to MB orbits where electrons, tunneling through at least eight gaps, encircle the entire diamond-shaped FS in the Z-R-A plane in the FBZ of \Z. The more sensitive torque data presented here allow us to see additional structures in the HF part of FFT around 17 and 21 kT arising from the first harmonic of A and B clusters, which to the best of our knowledge is another new result. Comparing the picture of MB orbits (Fig. \ref{fig_orbits}) with measured A and B clusters of peaks and their harmonics, it is confirmed that the QO frequency from the $\beta$ electron pocket in FS of \Z\ is $\beta=591(15)$ T, not 420 T as suggested in \cite{Z_2}.

FFT peaks corresponding to the individual $\alpha$ hole [$\alpha=241(4)$ T] and $\beta$ electron [$\beta=591(15)$ T] pockets are both visible also in the LF part of FFT, Fig. \ref{fig_LF}. Due to the dog-bone shape of the $\beta$ electron pocket a reasonable assumption is that there will be intraband tunneling through the neck of the $\beta$ pocket. We showed that this additional tunneling could well account for the observed prominent LF peaks $\gamma$ and $\eta$ with a tunneling path given in Fig. \ref{fig_orbits}. Also all other peaks in the LF part of FFT (which arise at fields above 20 T) can be interpreted as combined MB orbits that include one or more $\alpha$ pockets because every peak has an associated peak separated by $\alpha=241(4)$ T pocket, Fig. \ref{fig_LF}. 

The impact of changes in the Fermi energy and interlayer interaction on the FS of \Z\ (and other members of the same materials family) is still being intensively studied and the FFT spectra of QO in \Z\ is still not completely resolved \cite{Hu_2, Hosen, Z_1, H_1, Z_2}. In this work we present one possible scenario of electron dynamics in the MB regime in \Z\ which is supported by all our measurement results.

\subsection*{Acknowledgments}

This work was supported by Croatian Science Foundation under the project IP 2018 01 8912 and CeNIKS project cofinanced by the Croatian Government and the EU through the European Regional Development Fund - Competitiveness and Cohesion Operational Program (Grant No. KK.01.1.1.02.0013). Measurements were performed at High Field Magnet Laboratory (HFML) in Nijmegen. We acknowledge T. Klaser for XRD measurements. We thank J. R. Cooper for useful discussions.


\begin{thebibliography}{2}
\bibitem{synthesis}
Starting polycrystalline \Z\ was produced by closing a high purity elements in the vacuum sealed quartz ampule which is held at 1100\C\ for 4 days. Obtained polycrystalline material was characterized by powder XRD. Crystal growth took place in a  vacuum sealed quartz ampule with cca 1 g of starting polycrystalline material and 0.2 g of iodine placed in a temperature gradient of 1050-950\C\ for 21 days.
\bibitem{lever}
The crystal was glued to the end of the piezolever with its ab plane parallel to the flat surface of the lever. A dummy lever compensates background magnetoresistance signals using a Wheatstone bridge circuit driven by a floating 17 Hz current source. The voltage signal arising from the gravitational torque on the crystal when the sample stage is rotated in zero magnetic field gives the T-dependent sensitivity of the piezolever.
\end{thebibliography}
\end{document}